\documentclass[12pt]{article}
\usepackage{graphicx}
\usepackage{url}
\usepackage{cite}
\usepackage[margin=1.25 in]{geometry}
\usepackage[colorlinks = true, linkcolor = blue, urlcolor = blue,
      citecolor = blue, anchorcolor = blue]{hyperref}

\newcommand\pubnumber{SLAC-PUB-250501}
\newcommand\pubdate{May 2025}

\def\SLAC{SLAC National Accelerator Laboratory,
    Stanford University, Menlo Park, California 94025 USA}
\def\doeack{\footnote{Work supported by the US Department of Energy,
                     contract DE--AC02--76SF00515.}}

\def\Title#1{\begin{center} {\Large #1 } \end{center}}
\def\Author#1{\begin{center}{ \sc #1} \end{center}}

\def\submit#1{\begin{center}Submitted to {\sl #1} \end{center}}
\newcommand\pubblock{\rightline{\begin{tabular}{l}
        \pubnumber \\ \pubdate \end{tabular}}}
\newenvironment{Abstract}{\begin{quotation} \begin{center}
                       ABSTRACT
     \end{center}\bigskip  }{\end{quotation}}

\def\submit#1{\begin{center} #1\end{center}}
\def\Acknowledgements{\bigskip  \bigskip \begin{center} \begin{large}
             \bf ACKNOWLEDGEMENTS \end{large}\end{center}}
\def\beq{\begin{equation}}
\def\eeq#1{\label{#1}\end{equation}}
\def\eeqn{\end{equation}}

\newenvironment{Eqnarray}%
   {\arraycolsep 0.14em\begin{eqnarray}}{\end{eqnarray}}
\def\beqa{\begin{Eqnarray}}
\def\eeqa#1{\label{#1}\end{Eqnarray}}
\def\eeqan{\end{Eqnarray}}

\def\leqn#1{(\ref{#1})}

\let\bar=\overbar

\def\lsim{\mathrel{\raise.3ex\hbox{$<$\kern-.75em\lower1ex\hbox{$\sim$}}}}
\def\gsim{\mathrel{\raise.3ex\hbox{$>$\kern-.75em\lower1ex\hbox{$\sim$}}}}

\def\del{\partial}
\def\Dslash{\not{\hbox{\kern-4pt $D$}}}
\def\dslash{\not{\hbox{\kern-2pt $\del$}}}

\def\Dlr{\mathrel{\raise1.5ex\hbox{$\leftrightarrow$\kern-1em\lower1.5ex\hbox{$D$}}}}

\def\ee{e^+e^-}

\def\msb{{\bar{\scriptsize M \kern -1pt S}}}

\def\drb{{\bar{\scriptsize D \kern -1pt R}}}

\makeatletter
\def\section{\@startsection{section}{0}{\z@}{5.5ex plus .5ex minus
 1.5ex}{2.3ex plus .2ex}{\large\bf}}
\def\subsection{\@startsection{subsection}{1}{\z@}{3.5ex plus .5ex minus
 1.5ex}{1.3ex plus .2ex}{\normalsize\bf}}
\def\subsubsection{\@startsection{subsubsection}{2}{\z@}{-3.5ex plus
-1ex minus  -.2ex}{2.3ex plus .2ex}{\normalsize\sl}}

\renewcommand{\@makecaption}[2]{%
   \vskip 10pt
   \setbox\@tempboxa\hbox{\small #1: #2}
   \ifdim \wd\@tempboxa >\hsize     % IF longer than one line:
       \small #1: #2\par          %   THEN set as ordinary paragraph.
     \else                        %   ELSE  center.
       \hbox to\hsize{\hfil\box\@tempboxa\hfil}
   \fi}

\makeatother

\begin{document}
\begin{titlepage}
\pubblock

\vfill
\begin{center} 
\Title{What is the Hierarchy Problem?}
\vfill
\Author{
  Michael E. Peskin\doeack}
 \medskip
\SLAC 
\end{center}
\vfill
\begin{Abstract}
  Is there a Hierarchy Problem?  If so, what, exactly, is the problem?
  Almost every theorist has a personal answer to these questions.  In
  this article, I give my answer.   I will explain that the Hierarchy
  Problem is not a formal problem but rather our ignorance of a
  crucial physics explanation -- the explanation of the nature  of the
  Higgs boson.  Without the solution to this problem,
  we cannot make progress on the major questions of our field.
\end{Abstract} 
\vfill
\submit{To appear in a special issue of Nuclear Physics B : \\
``Clarifying Common Misconceptions in High Energy Physics and
Cosmology'', \\  Eoin O Colgain  and Yasaman Farzan, eds. }

\vfill
\end{titlepage}

\hbox to \hsize{\null}

%\newpage

\tableofcontents

\def\thefootnote{\fnsymbol{footnote}}
\newpage
\setcounter{page}{1}

\setcounter{footnote}{0}

\section{Introduction}

Today, the ``Hierarchy Problem'' is the question in particle physics
about which there is the least consensus.  There are many conceptions
of this problem and the related question of the ``naturalness'' of the
parameters of the Standard Model.   I doubt that --- absent any
illuminating experimental discovery --- a new theoretical paper could
meet the stated goal of this special issue of ``clarifying
common misconceptions''.    Still, I am glad to have this opportunity
to state my personal opinions on what the Hierarchy Problem really
asks and to offer my own route to a solution.   As to whether I am
dissolving confusion and simply adding to it, I leave to the reader to
judge.

The approach to the Hierarchy Problem in this paper is intentionally
narrow. In particular, I will only discuss ``the'' Hierarchy Problem,
the problem of the value of the Higgs mass term $\mu^2$ in the Standard
Model.   For those who seek a comprehensive review of the Hierarchy
Problem, including discussions of the Cosmological Constant Problem,
I strongly recommend the paper written by Nathaniel Craig for
the Snowmass 2021 study,
which attempts a complete survey of the literature~\cite{Craig:2022eqo}.

\section{Is there a Hierarchy Problem?}

The Hierarchy Problem in its simplest formulation concerns the
mechanism of the spontaneous breaking of electroweak $SU(2)\times
U(1)$ gauge symmetry in the Standard Model (SM).   I define the SM as
a renormalizable theory with exactly one fundamental Higgs scalar
doublet $\Phi$.  Under this set of assumptions, there is a unique
expression for the potential energy of the Higgs field,
\beq
V(\Phi) = \mu^2 |\Phi|^2 + \lambda |\Phi|^4 \ .
\eeq{treepot}
The parameters $\mu^2$ and $\lambda$ are  parameters of a
renormalizable quantum field theory and, as such, must be adjusted by
hand to values that fit the experimental data.   The observed spontaneous breaking
of electroweak symmetry (EWSB) requires $\mu^2 < 0$.

The SM with a single Higgs field has important virtues that must be
taken seriously.  In particular, it implies that lepton flavor
conservation and flavor universality is absolute, up to truly tiny
effects due to the neutrino masses, and it implies that flavor mixing
and CP violation in the hadronic weak decays is completely described
by the four Cabibbo-Kobayashi-Maskawa  (CKM) parameters.  It implies that
there is no flavor violation in Higgs boson couplings  and decays.  So
far, these restrictions hold up very well against experiment.   The SM
allows arbitrarily large mass hierarchies among the quarks and
leptons, and these hierarchies are stable under higher loop
corrections.

However, for those who would like to understand the reason for
EWSB, the SM is extremely frustrating.  In the SM, the parameter
$\mu^2$ is inserted by hand with no explanation.    In principle, the
origin, the value, and, especially, the sign
of $\mu^2$ ought to be found from an argument based
on physics.  There is a long history of models giving mechanisms of
EWSB, but every model requires new interactions in addition to those
in the SM.  The SM alone just throws up its hands.   The is the version of
the problem that I feel is the most important.   So, yes, there is
definitely a hierarchy problem.

Most discussions of the hierarchy problem are given in terms of the
instability of the value
or sign of $\mu^2$ under loop corrections. In the rest of this
section, I will describe this approach and comment on it.

Even at the 1-loop order, the radiative corrections to the $\mu^2$
parameter have a troubling form. One finds
\beq
         \mu^2 = \mu^2|_{\mbox{bare}}  -  {3 y_t^2\over
             8\pi^2}\Lambda^2  + {3
             \lambda\over 8 \pi^2}\Lambda^2       + {9 \alpha_w + 3
             \alpha'\over  16\pi} \Lambda^2  + \cdots \ .
           \eeq{murenorm}
  I have quoted this expression with the $\Lambda$ parameters
    being ultraviolet (UV) cutoffs on the Feynman integrals. These can
    have other interpretations, as I will discuss in the next
    paragraphs.   In all but the most naive models, these parameters
    take large dimensionful values corresponding to higher mass scales
    in nature.   The three $\Lambda$ parameters need not be
    identical.   Looking at the problem in this way, it seems that
    there are large positive and negative contributions to $\mu^2$
    that must be arranged to cancel to give the observed value $\mu^2
    \approx$ -(100 GeV)$^2$.   This is why a problem that might
    otherwise be called ``the $\mu^2$ ignorance problem'' is called
    the Hierarchy Problem.

    When they hear the Hierarchy Problem described in this way, many
    students are not concerned by this issue.  They point out that, 
     in dimensional regularization, the
     ultraviolet divergences disappear.  Isn't that a solution to
     this problem?  I find this explanation unacceptable. Let me first
     give a technical explanation.
     Dimensional regularization is a very convenient regulator
     because it sets integrals with no mass parameters equal to zero.
     In the case of the above diagrams, dimensional regularization
     renders them UV finite by subtracting an integral
     \beq
     \int {d^d k\over (2\pi)^d} {1\over k^2} = 0 \ .
     \eeq{subtractk}
     But one should also consider the physical meaning of this
     subtraction.
     The massless particles implied by \leqn{subtractk} do not exist.
    If any higher-mass particle  of mass $M$ contributes to the loop,
    it will, in dimensional regularization, give a contribution of the
    structure of one of the terms in \leqn{murenorm} with the
    $\Lambda^2$ replaced by $M^2$.  This includes particles associated
    with the completion of the SM to include quantum gravity.  The
    specific example of a contribution from   the heavy
    neutrino mass  $M_R^2$ in 
    the seesaw mechanism of neutrino mass
    generation is worked out in \cite{Vissani:1997ys}.
    Fluctuations of the $t$, $\Phi$, and $W$, $Z$ quantum fields at
    high momenta certainly do exist.  Even dimensional regularization
   tells us that these fluctuations contribute to the renormalization
    group $\beta$ function of the $\lambda$ parameter. In fact, they
    contribute to the prediction of the SM that its vacuum is
    unstable~\cite{Isidori:2001bm, Degrassi:2012ry}.  So their
    effect is not so easily dismissed~\cite{poletwo}.

    This technical argument comes back to my earlier displeasure with
    the SM.  Given that any modification of the SM leads to terms of the
    form of those in \leqn{murenorm}, the argument from dimensional
    regularization
    works only if we
    dismiss any correction to the SM, at any mass scale.   Then we are
    really giving up on the explanation of $\mu^2$.

    The Hierarchy Problem is often presented by quoting
    \leqn{murenorm} and then assuming that the $\Lambda$ parameters
    must be proportional to a very large mass scale such as the Planck
    scale.   This poses a striking problem, but it misses the
    essential physics question.  I will expand on this in the
    following section.

  Before going further, though, I would like to call attention to a
  benefit of taking the fluctuations of the top quark field at large
  momentum seriously.  Consider the vacuum of the top quark quantum
  field as represented by a Dirac sea.   It is obvious that generating
  a Higgs field vacuum expectation value that gives the top quark a
  mass opens a mass gap in the spectrum that  lowers the energy of the
  filled Dirac sea, which is the same as lowering the energy of the vacuum.
 I have found this argument helpful in clarifying  the
  mystery of EWSB for
  physics department colloquium audiences. Why not take it seriously?
  The idea that the
  largeness of the top quark Yukawa coupling relative to other
  dimensionless couplings of the SM is the driver of EWSB is very
  enticing. This was first  noticed in the context of
  supersymmetric grand unified
  theories~\cite{Ibanez:1982fr,Inoue:1982pi,Ellis:1982xz,Alvarez-Gaume:1983drc},
  but it applies in a very wide variety of explicit models of EWSB, as
  I will discuss below.

  \section{The Wilsonian point of view}

  Ken Wilson is often credited with the original statement requiring a
  natural explanation for the Higgs mass parameter, 
through his comment: ``It is interesting
to note that there are no weakly coupled scalar particles
in nature; scalar particles are the only kind of free
particles whose mass term does not break either an
internal or a gauge symmetry.''~\cite{Wilson:1970ag}.   However,
Wilson's ideas run through this problem at a deeper level than this.
At a time when particle theorists believed that hadron were
fundamental and rejected equations with pointlike couplings and UV
divergences,  Wilson held to the belief that
quantum field theories were not different from ordinary
quantum-mechanical systems.  He analyzed them by taking the UV cutoff
seriously as the largest energy scale  and working downward, rather
than considering the UV cutoff scale as an artifact to be removed by appropriate
incantations~\cite{Wilson:1965zzb}.  This led him to consider quantum
field theories as related to  models of phase transitions in
condensed mater and, eventually, to important discoveries about those systems.
  These discoveries doubled back to particle physics in the
concept of Wilsonian effective field theories, which now provide a
basic language for our  discussions of particle physics.

As a student of Wilson, I am thoroughly infected with that philosophy,
for better or worse.  Spontaneous breaking of symmetry is found in a
great variety of condensed matter systems --- magnets,
superconductors, liquid crystals, and more.  The condensed matter
context forces us to think about symmetry-breaking in these systems in
a different way.   Condensed
matter systems are made of atoms, and the laws of atomic physics have
been fixed  since discovery of quantum mechanics.  This means that the
explanations for
these examples of
spontaneous symmetry breaking must be derivable from those laws.  This
makes the
study of phase transitions in condensed matter a fascinating
subject, with many surprising mechanisms leading to the observed nontrivial ground
states~\cite{ChaikinLubensky}.   Thinking in this way gets us away from
simplifying the hierarchy program to the explanation of 
a large ratio of  mass scales.

Then, {\it  the real
goal of our  pursuit of the origin of $\mu^2$ should be to find a compelling
physical mechanism that explains the origin of EWSB.}   This mechanism
might be one already known in condensed matter physics, or it might be 
one that is 
completely new.  In any case, that physics is not present in the SM, a
theory that is completely weak-coupling at short distances.
{\it This point of view requires that there
must be new fundamental interactions working at shorter distances than
the ones that we have probed so far.}   That gives an important goal
for further exploration in high-energy physics.

An example of a physical mechanism leading to spontaneous symmetry breaking
in particle physics is the explanation of chiral
symmetry breaking in the strong interactions. From the development of
the theory of weak interactions in the 1950's, it was understood that
the strong interactions should be invariant under chiral symmetries,
even though QCD,  which makes this statement obvious, was many years
in the future.  In 1961, Nambu and
Jona-Lasinio proposed that the chiral symmetry of the strong
interactions could be spontaneously broken by same mechanism that
causes superconductivity -- fermion pair
condensation~\cite{Nambu:1961tp}.  Throughout the 1960's, this
idea and the related idea of the quark model were considered to be
hopelessly naive.  Progress was made through the language of ``current
algebra''~\cite{Gell-Mann:1964hhf} in papers that are very difficult
to read today.
Finally, these physically transparent ideas found their place within
the theory of QCD with asymptotic freedom.

My goal for the hierarchy problem is similar.   Can we find a
cogent explanation for EWSB based on an explicit physical model?  This
does not need to be an ultimate explanation or one correct all the way
to the Planck scale.    We can build our  model of fundamental physics
step by step.

This is the step that we need to take now.  Different models of EWSB
lead to different models of the Higgs boson, for example, as a member
of a supersymmetry multiplet, as member of a global symmetry
multiplet, as  a composite particle.   Each hypothesis leads to
distinct theories of the most important issues in elementary particle
physics -- the origin of the hierarchy of fermion masses and mixings,
the origin of the baryon asymmetry in nature, the origin of neutrino
masses. Many hypotheses for the problem of the origin of dark matter
also depend of this choice.   Without the solution to the Hierarchy Problem, as I have
framed it here, it will be difficult to find the correct path to
address any of these questions, much less to find the correct model.

\section{The Hierarchy Problem before the LHC Results}

I saw the Hierarchy Problem in this way also before the start of
the LHC experimental program.   At that time, I was optimistic that we would
obtain clues to this problem from the discovery of new particles at the
LHC.

For many members of our community, the idea that drove the
search for new particles was the beauty of supersymmetry.
Supersymmetry is,  after all, a unique extension of Poincar\'e
invariance.  It is a central element of string theory and of most   
successful theories of grand unification.  Supersymmetry also provides
a natural mechanism for cutting off the quadratic divergences in
\leqn{murenorm},  since the quadratic divergences in scalar masses
parameters must cancel in a completely supersymmetric theory.  And, if
the Higgs boson should be found at a mass below 1~TeV, its
superpartners should also be found there.

For me, though, supersymmetry had a different attraction.
Supersymmetry provides a quantitative physical mechanism by which the top quark
Yukawa coupling can provide a physical mechanism for
EWSB.  In
supersymmetric theories, the logarithmic renormalization of soft supersymmetry
breaking scalar mass terms by Yukawa couplings is negative.  So, if
supersymmetry is spontaneously broken, as it must be to provide a
realistic theory, then the renormalization of the symmetry-breaking
mass terms will lead to a vacuum instability for one or more scalar
fields.  There are many, many new scalar fields in a supersymmetric
extension of the SM, leading to many possibilities for symmetry
breaking.
But, due to the large size of the top quark
Yukawa coupling and some convenient group theory factors,
it is the Higgs doublet scalars for which
this effect is largest, leading to exactly the pattern of EWSB
found in the
SM~\cite{Ibanez:1982fr,Inoue:1982pi,Ellis:1982xz,Alvarez-Gaume:1983drc}.
This mechanism is highlighted in my 2006 TASI lectures on
supersymmetry~\cite{Peskin:2008nw}.

If we give up the idea that the solution to EWSB must also provide a
fundamental theory up to very high energies, then there are many 
models that use the largeness of the top quark Yukawa coupling to
drive other mechanisms for EWSB.  In 1984, Kaplan and Georgi
introduced the idea that the Higgs boson  could be a composite
particle bound by a new set of strong
interactions~\cite{Kaplan:1983fs}.   The Higgs boson  could be light
compared to the symmetry-breaking scale of the strong interaction
theory if it were a Goldstone boson of a dynamically broken global symmetry
of that theory.   The Higgs boson mass term would be supplied by
radiative corrections induced by terms in the theory that did not
respect that global symmetry.  Later, this mechanism would be
explicitly realized (and some difficulties of the Kaplan-Georgi model
solved) in Little Higgs
theories~\cite{Arkani-Hamed:2002ikv, Schmaltz:2005ky}.   In these
models, a vector-like fermion top quark partner would cancel the
quadratic divergences of the top quark loop correction to $\mu^2$,
leaving over a naturally negative contribution.

A similar mechanism was found to work in models with an extra space
dimension.   In these models, the Higgs boson doublet is identified
with the 5th component of a set of gauge fields in the higher
dimension.  This identification leads to the cancellation of quadratic
divergences.    The contribution to the Higgs mass term from the top
quark is again negative, but the contributions from the Kaluza-Klein
resonances of the 5-dimensional top quark field are positive and cut
off the ultraviolet divergence.   Here also, radiative corrections due
to the top quark Yukawa coupling lead to a 
naturally negative value of $\mu^2$ ~\cite{Contino:2003ve}.

In all of these cases, new particles are needed to build a complete
model of EWSB --- the supersymmetric partners of the top quark in the
case of supersymmetry, vectorlike fermion top quark partners in the
cases of nonsupersymmetric models.    These particles can be heavier
than the Higgs boson, but fine-tuning is needed to push their masses
above 1~TeV.   In the case of supersymmetry, where the entire theory
is described by weak-coupling interactions, a rather sophisticated
literature on fine-tuning developed, suggesting quite
strong  upper limits on the masses of these
particles~\cite{Barbieri:1987fn,Feng:2013pwa}.  In the
non-supersymmetric cases, the limits are weaker but still in the
region of  1-3~TeV.

So far, none of these new particles has been discovered at the LHC.
Especially because of the stringent expectations for supersymmetry,
most members of our community have come to the opinion that these
particles do
not exist. Still, especially for  the vectorlike fermions,  there is
opportunity to extend the searches and
discover these particles at the
HL-LHC. I hope that LHC experimenters will take this opportunity seriously.

Many theorists now are trying to explain that they never actually
predicted the discovery of new particles at the LHC, or that they gave
reasons why these potential discoveries were already excluded.  I am not one of
them.  I feel that the models that I have reviewed in this section remain
compelling  ideas for the mechanism of EWSB.   Even if
nature does not choose models of these types, we ought to own the
reasoning that led to them. 
The true explanation for EWSB might be very close, if only some
further new idea can be added to the mix.

\section{The ``Post-Naturalness Era''}

The failure of the LHC experiments to discover new particles has been
discouraging for many  theorists.   This has led to a search for a
solution to the Hierarchy Problem outside of the realm of particle
physics model-building.   Suggestions include the use of the Anthropic
Principle, mechanisms in the physics of the early universe,  and
possible UV/IR connections in the quantum theory of gravity.   Many of
these models are explained in  Anson Hook's review
paper~\cite{Hook:2023yzd},
which specifically
concerns  nontrivial
mechanisms for relieving the  fine-tuning of $\mu^2$ in models of
fundamental scalar fields.

Excuse me that I am very cool to these ideas.   Particle theorists
often view  scalars as part of the general equipment of nature, to be
added at will to any theory.  As I have already stated above, I
believe that scalar fields in nature should have a purpose.  If they
obtain  vacuum expectation values, those expectation values should be
tied to the masses of particles or to another dimensionful
reference point.   It is possible that the vacuum expectation value of the
Higgs field could be chosen randomly.  But it would be much more
pleasing to explain this value in an underly physical picture.

In a 2017 paper, Gian Giudice presented this  direction of research in
a very optimistic way~\cite{Giudice:2017pzm}. In an article titled
``The Dawn of the
Post-Naturalness Era'',  he describes the exclusion by the LHC of the
models described in the previous section as a true crisis in theoretical
particle physics, a crisis that would, in his view, be resolved only
by the invention of revolutionary new ideas.  Particle theorists are
eager for revolutions, and also for ``final theories'' that fill
the gap between currently explored energies and the Planck scale.  It
is bold to try to imagine these new ideas, but I feel it is not yet
necessary.  Perhaps conventional quantum field theory has still  not shown us all of its
tricks.

\section{Three Hierarchy Problems}

In this more pragmatic way of thinking, it is useful to divide the
traditional Hierarchly Problem into three problems.

The first is the traditional ``Hierarchy Problem'':  Why does the physics
of electroweak symmetry breaking occur at energies so much lower than
the Planck scale?   There are many possible answers to this question.
Supersymmetry with a low scale of its spontaneous breaking is one.
Asymptotic freedom, for which a small coupling constant at an original
scale leads to a
dynamical generation of a new scale that is exponentially smaller, is
another.  Actually,  we know that this is the explanation for the
hierarchy problem of QCD, explaining why the mass of the proton is so
much smaller than the Planck scale.

To formulate a solution to this problem, we first need to know what
are the basic particles and fields that cause EWSB.   Because  new
particles have not been discovered at the LHC,  the nature of these
fundamental particles is still hidden from us.

The second is what one might call the ``Problem of Scalars''.  This is
the flip side of the idea that one should add scalars to a model to
give it more interesting dynamics.   Scalars have dimension 1 and can
couple to other fields almost without restriction.  Even extending the
Higgs sector of the SM to two Higgs doublets  extends the parameter
set to  4 mass parameters and  10 quartic  parameters, counting possibly
complex coefficients as two real parameters (reducible to a total of
11 after field redefinitions).

In the SM, the masses of quarks and leptons come from Yukawa
interactions of the of these fermions with Higgs field.
These couplings are given in terms of 3 complex $3\times 3$
matrices, a total of  54 fundamental parameters.   Of these, only 14
-- the nine quark and lepton masses, the 4 CKM parameters, and the
$\theta$
parameter, are potentially observable.  The rest can be removed by
field redefinitions.  This leads to enormous ambiguity in the search
for a theory of fermion masses and mixings.  The history of theories
of flavor reflects this.  In 1977, Harold Fritzsch gave a
model~\cite{Fritzsch:1977za}
that successfully predicted the relation
\beq
\tan \theta_c \approx \sqrt{m_d/m_s}  \ .
\eeqn
It has been all downhill since then.  Despite the efforts of many of
the leading theorists of my
generation --- including Jogesh Pati, Savas Dimopoulos,
Lawrence Hall, Graham Ross, and Yossi Nir --- no one would say today,
``Give me one more decimal place on the CKM parameters and this
problem will be solved.''   The explanation for the neutrino
masses, which involve a $3\times 3$ complex seesaw mass matrix in
additional to a Yukawa matrix, is even further away.

In addressing this problem, it would be tremendously illuminating to
discover the organizing principle for the scalar couplings.   Do they
come from compositeness of  the Higgs boson?  from the mixing of the
lightest Higgs field with heavier partners?   from a hierarchy of
couplings generated by perturbation theory?  An aspect of quantum
field theory that we know little about today is the dynamics and
possible  composite particles of chiral gauge theories.   Maybe
this is a place that we will find clues to these questions.

It would be wonderful to learn the correct explanation from
experiment.  But, again, our ignorance of the fundamental nature of
the Higgs field and our lack of clues about what stands behind the SM
is an impediment here.

The third problem is the ``Little Hierarchy Problem''.    This is the
question of why the masses of new particles needed for a dynamical model of EWSB
are so much heavier than the mass of Higgs boson.   In the context of
a search for the dynamical explanation of EWSB, this is a constraint
on the possible answers: They must have some feature that leads to an
unanticipated mass gap between the $\mu^2$ parameter and the masses of
the lightest new particles.

In my opinion, the Little Hierarchy Problem is the one that offers the
best chance for a solution now.  Only a small hierarchy is needed.
In addition,  we need the answer to this
question to make progress on either of the previous two hierarchy
problems, which cannot even be properly posed without new information
about the nature of the new particles to be found at energies above
those currently probed by the LHC.

\section{Solutions to the Little Hierarchy Problem}

There are solutions to the Little Hierarchy Problem in the literature,
but no one is very impressed with them. In this section, I will
discuss three of these. In all three cases, the next set of
fundamental interactions beyond the SM has two levels, one immediately
generating the Higgs potential and another, at a higher mass scale,
being
the more fundamental cause of the symmetry-breaking.  I will refer to
these as the intermediate and the fundamental mass scales.
In their current state, these
models seem artificial and overly complex.   However, they provide
starting points.  

The first idea is to include more fields in the region between the
intermediate and fundamental scales in the model.   There are many
examples of models that use this strategy implicitly.    The strategy
is made more
explicit in work of mine with Yoon, where we call it
``competing  forces''~\cite{Yoon:2017cty}.   In the example we
discuss, the chiral top quark naturally generates a term in the Higgs potential
with a negative $\mu^2$. In the same model, an additional vector-like
fermion generates a term in the Higgs potential with a positive
$\mu^2$.  Playing these two effects against one another give a Higgs
potential that requires only a 
modest level of tuning to move new particles out of the range
explored by the LHC.   The new competing fermion might have its own
reason for existence, for example, to generate the cosmic dark matter.

The second idea, ``2-stage symmetry breaking'' takes advantage of a
property of models in which the fundamental symmetry breaking is due
to fermion condensation, with the Higgs boson identified with a
Goldstone boson created in this process. 
In such models, mass generation for the Goldstone boson
is forbidden by two different chiral symmetries, one associated with
the left-handed fundamental fermions, the other with the right-handed
fundamental fermions.   Breaking both symmetries can require two
insertions of weakly coupled chiral symmetry breaking perturbations,
leading to a formula for the induced $\mu^2$ term of the form
\beq
\mu^2 =  - {3 \alpha_w y_t^2 \over 8\pi^2} f^2 \ ,
\eeqn
with $f$ the mass scale of the fundamental level; often, this is
enhanced by a factor $\log f^2/m_t^2$.   This 2-stage generation of
$\mu^2$ is a property of the Littlest Higgs
model~\cite{Arkani-Hamed:2002ikv} and is analyzed further in
\cite{Perelstein:2003wd}.
The model does require a new vectorlike top partner at the
intermediate scale, which perhaps could be generated as a massless
composite fermion of the fundamental-level theory.

The third idea is one special to supersymmetry, ``Dirac
gauginos''~\cite{Fox:2002bu}.   In this strategy, the gaugino sector
is $N=2$ supersymmetric, while the matter sector has $N=1$
supersymmetry.  Supersymmetry can be broken by a D term in the N=2
sector.  In the matter sector, this generates supersymmetry-breaking
mass terms that are ``supersoft'', that is, very insensitive to
details of the fundamental scale.   The $\mu^2$ term is generated both
directly, by an electroweak supersoft term, and in two stages, using
the top quark Yukawa coupling. Schematically
\beq
  \mu^2 \approx  {\alpha_w M_1^2 \over \pi }-  {3\lambda_t^2 m_{\tilde
        q}^2\over 4\pi^2} \log {M_3\over m_{\tilde q}} \ ,
      \eeqn
where $M_a$ are gaugino masses and $m_{\tilde q}$ is a squark mass.
Effectively, this 
mechanism automatically uses both of the above strategies to lower
$\mu^2$ relative to the fundamental supersymmetry breaking mass scale.
Some further development of this idea relevant to the hierarchy
problem can be found in \cite{Baer:2015rja, Chakraborty:2018izc,Cohen:2020ohi}.

There is a ample room to improve these strategies, and perhaps
find new ones, in pursuit of an elegant model that solves the Little
Hierarchy problem.  This direction deserves more attention
from the particle theory community.

\section{The Hierarchy Problem and future high-energy colliders}

Particle theorists often confine their thinking to their own narrow
domain.  However, our beliefs about the Hierarchy Problem have
important implications for experimental particle physics that I feel
cannot be ignored.

In the development of the SM, the progress of theory relied strongly
on surprises from experiment.  The $\tau$-$\theta$ puzzle of kaon
decays opened the path to an understanding of the weak interactions.
The SLAC-MIT deep inelastic scattering experiments and the discovery
of the $J/\psi$ shattered the
notion that hadrons were fundamental particles and opened the door to
the quark model and QCD. Even after the SM was formulated and began to
be tested with precision, the heaviness of the top quark was a new
discovery that now informs our ideas about beyond-SM physics.   We need
new experimental surprises to guide further progress.

For the next high energy physics collider, there is a general
consensus and a well-developed
physics case for an $\ee$ Higgs factory~\cite{Dawson:2022zbb}.
This will make precision measurements that may point to the next
energy scale.   But in order to actually reach that scale, we will need a collider at
energies substantially higher than those of the LHC.    There is
much interest now in planning for a 100~TeV proton collider,
or, more generally, for a ``10~TeV pCM'' (parton CM energy)
collider~\cite{P5:2023wyd}.

The cost of any 10~TeV pCM collider will be in the \$10 B range.  A
less expensive machine might be justified by the  importance of exploring
for new fundamental interactions beyond our current knowledge.   For
particle physicists, the importance of this goal is obvious.   But for
our colleagues in other fields of science, and for our government
sponsors, this is much less clear. Already, we are hearing from many
sources that, with the SM complete, particle physics is finished, and that
the funding it requires for a next step  is better spent in other areas.  Especially at
this level of cost, we will be
asked what, more precisely, we expect to discover, and what energy is
actually needed to achieve that goal.   Our physics arguments for a
higher energy collider need to be much stronger.

Because the SM can be extrapolated to the Planck scale, there is no
guaranteed discovery at such a machine (as there was, for example, for
the LHC).   Still, we can make a strong argument for a 10 TeV pCM
collider if we can argue that this collider gives us an {\it
  opportunity} to discover and characterize a new fundamental
interaction of nature.

Do you believe that this is so?   The mechanistic view of the
Hierarchy Problem that I have presented in this paper leads to a very
different answer from models that simply motivate a hierarchy of
unknown size from the randomness of nature or from the influence of
constraints from gravity or cosmology. In those models, higher energy
scales in physics may be very far away, or, even, might not be needed
at all
below the scale of gravity.    Mechanistic models of the
Hierarchy Problem are different in this respect.  They
call for new particles at energy scales that are
higher than those currently studied but are linked to the weak interaction scale.

I feel strongly the development of new mechanistic models of EWSB is essential
 if we wish to advocate for higher-energy colliders.
Can we claim that there is an opportunity to discover new
fundamental interactions if we can obtain another factor of 10 in
collision energy?
Can we illustrate this with compelling models that address EWSB and
other major
questions of particle physics?   Can we make a quantitative argument
that 10~TeV is a important milestone?   If we cannot answer these
questions for ourselves, we will not be able to persuade others.

Today, 10~TeV pCM colliders are still out of reach. In the next
decades, the particle physics community must put serious effort into
the development of new tools and technologies to reach those
energies.  In parallel, theorists must work to define the
experimental program that these  accelerators will carry out. We
cannot do this without confronting the Hierarchy Problem and bringing
new ideas to its solution.  This
is a responsibility that the particle theory community must
address.

\Acknowledgements

The ideas expressed in this paper were shaped by many conversations,
but I would particularly wish to thank Ann Nelson.  Ann was a skeptic
of the simple models discussed in Section 4, but a strong believer in
the importance of model-building to find the way forward.  I thank
Nathaniel Craig and Yasaman Farzan for useful comments on the manuscript.

This work was supported by the US Department of Energy under
                     contract DE--AC02--76SF00515.
I am grateful to Nathaniel Craig and David Gross for a visit to 
the Kavli Institute for  Theoretical Physics that allowed me 
to preview these ideas before a skeptical and discerning audience.
That visit was supported by National Science Foundation grant NSF
PHY-2309135 to the KITP.


\begin{thebibliography}{99}

\bibitem{Craig:2022eqo}
N.~Craig,
``Naturalness: Past, Present, and Future,''
Eur. Phys. J. C \textbf{83}, 825 (2023)
%doi:10.1140/epjc/s10052-023-11928-7
[arXiv:2205.05708 [hep-ph]].

\bibitem{Vissani:1997ys}
F.~Vissani,
``Do experiments suggest a hierarchy problem?,''
Phys. Rev. D \textbf{57}, 7027 (1998)
%doi:10.1103/PhysRevD.57.7027
[arXiv:hep-ph/9709409 [hep-ph]].  I thank Howard Baer for this reference.


\bibitem{Isidori:2001bm}
G.~Isidori, G.~Ridolfi and A.~Strumia,
``On the Metastability of the Standard Model Vacuum,''
Nucl. Phys. B \textbf{609}, 387-409 (2001)
%doi:10.1016/S0550-3213(01)00302-9
[arXiv:hep-ph/0104016 [hep-ph]].

\bibitem{Degrassi:2012ry}
G.~Degrassi, S.~Di Vita, J.~Elias-Miro, J.~R.~Espinosa, G.~F.~Giudice, G.~Isidori and A.~Strumia,
``Higgs Mass and Vacuum Stability in the Standard Model at NNLO,''
JHEP \textbf{08}, 098 (2012)
%doi:10.1007/JHEP08(2012)098
[arXiv:1205.6497 [hep-ph]].

\bibitem{poletwo}
  A way to see the presence of these fluctuations within dimensional
  regularization is to note that they create a nonzero pole in the
  Higgs field propagator at  $d=2$.  This contrasts with the situation
  of the electromagnetic vacuum polarization.   In that case, the
  quadratic divergence is required to vanish by QED gauge invariance.
  Dimensional regularization respects gauge invariance, and so it not
  only removes the divergence at $d=4$ but also it removes the pole at
  $d=2$ so that the QED Ward-Takahashi identity is valid for all
  $d$.   See M. E. Peskin and D. V. Schroeder, {\it An Introduction to
    Quantum Field Theory} (CRC Press, 1995), Section 7.5.

  \bibitem{Ibanez:1982fr}
L.~E.~Ibanez and G.~G.~Ross,
``SU(2)-L x U(1) Symmetry Breaking as a Radiative
Effect of Supersymmetry Breaking in Guts,''
Phys. Lett. B \textbf{110}, 215-220 (1982).
%doi:10.1016/0370-2693(82)91239-4


\bibitem{Inoue:1982pi}
K.~Inoue, A.~Kakuto, H.~Komatsu and S.~Takeshita,
``Aspects of Grand Unified Models with Softly Broken Supersymmetry,''
Prog. Theor. Phys. \textbf{68}, 927 (1982)
[erratum: Prog. Theor. Phys. \textbf{70}, 330 (1983)].


\bibitem{Ellis:1982xz}
J.~R.~Ellis, L.~E.~Ibanez and G.~G.~Ross,
``Supersymmetric Grand Unification,''
Nucl. Phys. B \textbf{221}, 29-67 (1983).
%doi:10.1016/0550-3213(83)90618-1

\bibitem{Alvarez-Gaume:1983drc}
L.~Alvarez-Gaume, J.~Polchinski and M.~B.~Wise,
``Minimal Low-Energy Supergravity,''
Nucl. Phys. B \textbf{221}, 495 (1983).
%doi:10.1016/0550-3213(83)90591-6


\bibitem{Wilson:1970ag}
K.~G.~Wilson,
``The Renormalization Group and Strong Interactions,''
Phys. Rev. D \textbf{3}, 1818 (1971).
%doi:10.1103/PhysRevD.3.1818

\bibitem{Wilson:1965zzb}
K.~G.~Wilson,
``Model Hamiltonians for Local Quantum Field Theory,''
Phys. Rev. \textbf{140}, B445 (1965).
%doi:10.1103/PhysRev.140.B445

\bibitem{ChaikinLubensky}
  P. M. Chaikin and T. C. Lubensky,
  {\it Principles of Condensed Matter Physics}
  (Cambridge University Press, 1995).

\bibitem{Nambu:1961tp}
Y.~Nambu and G.~Jona-Lasinio,
``Dynamical Model of Elementary Particles Based on an Analogy
with Superconductivity. 1,''
Phys. Rev. \textbf{122}, 345 (1961).
%doi:10.1103/PhysRev.122.345

\bibitem{Gell-Mann:1964hhf}
M.~Gell-Mann,
``The Symmetry Group of Vector and Axial Vector Currents,''
Physics Physique Fizika \textbf{1}, 63-75 (1964).
%doi:10.1103/PhysicsPhysiqueFizika.1.63

\bibitem{Peskin:2008nw}
M.~E.~Peskin,
``Supersymmetry in Elementary Particle Physics,''
[arXiv:0801.1928 [hep-ph]].


\bibitem{Kaplan:1983fs}
D.~B.~Kaplan and H.~Georgi,
``SU(2) x U(1) Breaking by Vacuum Misalignment,''
Phys. Lett. B \textbf{136}, 183 (1984).
%doi:10.1016/0370-2693(84)91177-8

\bibitem{Arkani-Hamed:2002ikv}
N.~Arkani-Hamed, A.~G.~Cohen, E.~Katz and A.~E.~Nelson,
``The Littlest Higgs,''
JHEP \textbf{07}, 034 (2002)
%doi:10.1088/1126-6708/2002/07/034
[arXiv:hep-ph/0206021 [hep-ph]].

\bibitem{Schmaltz:2005ky}
M.~Schmaltz and D.~Tucker-Smith,
``Little Higgs review,''
Ann. Rev. Nucl. Part. Sci. \textbf{55}, 229 (2005)
%doi:10.1146/annurev.nucl.55.090704.151502
[arXiv:hep-ph/0502182 [hep-ph]].

\bibitem{Contino:2003ve}
R.~Contino, Y.~Nomura and A.~Pomarol,
``Higgs as a Holographic Pseudo Goldstone Boson,''
Nucl. Phys. B \textbf{671}, 148 (2003)
%doi:10.1016/j.nuclphysb.2003.08.027
[arXiv:hep-ph/0306259 [hep-ph]].

\bibitem{Barbieri:1987fn}
R.~Barbieri and G.~F.~Giudice,
``Upper Bounds on Supersymmetric Particle Masses,''
Nucl. Phys. B \textbf{306}, 63 (1988).
%doi:10.1016/0550-3213(88)90171-X


\bibitem{Feng:2013pwa}
J.~L.~Feng,
``Naturalness and the Status of Supersymmetry,''
Ann. Rev. Nucl. Part. Sci. \textbf{63}, 351 (2013)
%doi:10.1146/annurev-nucl-102010-130447
[arXiv:1302.6587 [hep-ph]].


\bibitem{Hook:2023yzd}
A.~Hook,
``New Solutions to the Gauge Hierarchy Problem,''
Ann. Rev. Nucl. Part. Sci. \textbf{73}, 23 (2023).
%doi:10.1146/annurev-nucl-102422-080830


\bibitem{Giudice:2017pzm}
G.~F.~Giudice,
``The Dawn of the Post-Naturalness Era,''
%doi:10.1142/9789813238053\_0013
[arXiv:1710.07663 [physics.hist-ph]].

\bibitem{Fritzsch:1977za}
H.~Fritzsch,
``Calculating the Cabibbo Angle,''
Phys. Lett. B \textbf{70}, 436-440 (1977).
% doi:10.1016/0370-2693(77)90408-7

\bibitem{Yoon:2017cty}
J.~Yoon and M.~E.~Peskin,
``Competing Forces in Five-Dimensional Fermion Condensation,''
Phys. Rev. D \textbf{96}, 115030 (2017)
%doi:10.1103/PhysRevD.96.115030
[arXiv:1709.07909 [hep-ph]]; ``Dissection of an $SO(5) \times U(1)$
Gauge-Higgs Unification Model,''
Phys. Rev. D \textbf{100}, 015001 (2019)
%doi:10.1103/PhysRevD.100.015001
[arXiv:1810.12352 [hep-ph]].

\bibitem{Perelstein:2003wd}
M.~Perelstein, M.~E.~Peskin and A.~Pierce,
``Top Quarks and Electroweak Symmetry Breaking in Little Higgs Models,''
Phys. Rev. D \textbf{69}, 075002 (2004)
%doi:10.1103/PhysRevD.69.075002
[arXiv:hep-ph/0310039 [hep-ph]].

\bibitem{Fox:2002bu}
P.~J.~Fox, A.~E.~Nelson and N.~Weiner,
``Dirac Gaugino Masses and Supersoft Supersymmetry Breaking,''
JHEP \textbf{08}, 035 (2002)
%doi:10.1088/1126-6708/2002/08/035
[arXiv:hep-ph/0206096 [hep-ph]].

\bibitem{Baer:2015rja}
H.~Baer, V.~Barger and M.~Savoy,
``Upper Bounds on Sparticle Masses from Naturalness or How to
Disprove Weak Scale Supersymmetry,''
Phys. Rev. D \textbf{93}, 035016 (2016)
%doi:10.1103/PhysRevD.93.035016
[arXiv:1509.02929 [hep-ph]].

\bibitem{Chakraborty:2018izc}
S.~Chakraborty, A.~Martin and T.~S.~Roy,
``Charting Generalized Supersoft Supersymmetry,''
JHEP \textbf{05}, 176 (2018)
%doi:10.1007/JHEP05(2018)176
[arXiv:1802.03411 [hep-ph]].

\bibitem{Cohen:2020ohi}
T.~Cohen, N.~Craig, S.~Koren, M.~Mccullough and J.~Tooby-Smith,
``Supersoft Top Squarks,''
Phys. Rev. Lett. \textbf{125}, 151801 (2020)
%doi:10.1103/PhysRevLett.125.151801
[arXiv:2002.12630 [hep-ph]].


\bibitem{Dawson:2022zbb}
S.~Dawson, \textit{et al.}
``Report of the Topical Group on Higgs Physics for Snowmass 2021:
The Case for Precision Higgs Physics,''
[arXiv:2209.07510 [hep-ph]].

\bibitem{P5:2023wyd}
S.~Asai \textit{et al.} [P5 Panel],
``Exploring the Quantum Universe: Pathways to Innovation
and Discovery in Particle Physics,''
%doi:10.2172/2368847
[arXiv:2407.19176 [hep-ex]].


\end{thebibliography}
\end{document}